\documentclass[aps,prl,reprint,superscriptaddress,amsmath,floatfix]{revtex4-2}

\usepackage{graphicx}
\usepackage[colorlinks=true,linkcolor=black,citecolor=blue,urlcolor=blue]{hyperref}
\usepackage{dcolumn}
\newcommand{\vc}[1]{\boldsymbol{#1}} 

\begin{document}
\title{Triplon-mediated pairing and the superconducting gap structure in bilayer nickelates}
\author{Huimei Liu}
\affiliation{National Laboratory of Solid State Microstructures and School of Physics, Nanjing University, Nanjing 210093, China}
\affiliation{Collaborative Innovation Center of Advanced Microstructures, Nanjing University, Nanjing 210093, China}
\author{Giniyat Khaliullin}
\affiliation{Max Planck Institute for Solid State Research, Heisenbergstrasse 1, D-70569 Stuttgart, Germany}

\begin{abstract}
We investigate the superconducting gap structure in bilayer nickelates within a model in which conduction bands of $d_{x^2\!-\!y^2}$ symmetry coexist with localized $d_{3z^2\!-\!r^2}$ spins. Strong interlayer coupling drives the local moments into a singlet ground state, whose virtual singlet–triplet excitations (“triplons”) mediate the pairing interaction. This mechanism yields interband $s_{\pm}$ pairing, with opposite signs of the order parameter on the two ($\alpha$ and $\beta$) bands. The calculated tunneling spectra reproduce the recently observed two-peak structure, the larger gap on the $\alpha$ band despite its smaller density of states, and the gap anisotropy. The results provide strong evidence for the triplon-mediated pairing mechanism in bilayer nickelates.

\end{abstract}

\date{\today}
\maketitle

The recent discovery of superconductivity in bilayer nickelates under high pressure and epitaxial strain has opened a new frontier in the study of unconventional high-$T_c$ superconductors~\cite{Sun23a,Wan24,Ko25,Zho25,Liu25,Pup25}. Different from cuprates, where superconductivity emerges from a single half-filled $d_{x^2\!-\!y^2}$ orbital, bilayer nickelates involve a mixed-valence Ni$^{2.5+}$ state with active $d_{x^2\!-\!y^2}$ and $d_{3z^2\!-\!r^2}$ orbitals, fundamentally altering the nature of low-energy correlations and pairing. The interplay of orbital physics, strong interlayer coupling, electron correlations, and lattice effects creates a complex theoretical landscape, and the nature of pairing mechanisms in nickelates remains under active debate (see, e.g.,~\cite{Luo23,Yan23a,Yan23b,Qin23,Liu23,Oh23,Lia23,Lec23,Qu23,Jia25,Sak24,Che24,Lu24,Yi25,Kha26,Qiu25,Wat26,Wang25} and references therein).

Recent experiments have reported the superconducting gaps in bilayer nickelates~\cite{Fan25,She25,Sun25b,Guo25,Cao25,Wan26,Lia26}. 
The distinct two-peak structure is observed in the tunneling spectra~\cite{Fan25,Lia26}, which can be associated with the  pairing gaps in the bonding and antibonding bands of a bilayer lattice. It is particularly striking that the band with the smaller density of states at the Fermi level hosts the larger superconducting gap, and both gaps display angular anisotropy. These observations impose strong constraints on the theoretical models and pairing mechanisms.

In this Letter, we  present a theory that accounts for these key observations. Building on the earlier proposal that interlayer singlet-triplet excitations of the localized $d_{3z^2\!-\!r^2}$ electrons can mediate the pairing interaction~\cite{Kha26}, we develop a microscopic model which quantifies the superconducting gap structure in bilayer nickelates. Including orbital hybridization effects, we derive the bilayer-split electronic bands. We then obtain the effective interactions between localized spins and itinerant electrons arising from Hund’s coupling and intersite Kondo exchange. The resulting BCS Hamiltonian exhibits a distinct spin and band-flavor structure, leading to interband $s_{\pm}$ pairing with opposite signs and different amplitudes of the order parameter on the bonding and antibonding bands. The gaps are anisotropic in momentum space due to Kondo exchange contributions to pairing. We present the superconducting tunneling spectra predicted by our model. The theory reproduces well the observed two-peak structure and gap hierarchy, lending strong support to the triplon-mediated pairing mechanism in this new family of unconventional superconductors.

{\it The model: spin singlets, triplons, and conduction bands}.---In bilayer nickelates, the nominal valence of Ni is close to Ni$^{2.5+}$, corresponding to an average $d^{7.5}$ configuration. The partially filled $e_g$ orbital doublet contains $\sim 1.5$ electrons per Ni. The tetragonal crystal field splits the doublet into a lower $d_{3z^2\!-\!r^2}$ and a higher $d_{x^2\!-\!y^2}$ orbital levels. For a sufficiently large splitting, of order $1$~eV, as realized, e.g., in $ab$-plane compressed superconducting films~\cite{Bha25,Yi25}, one electron per Ni predominantly occupies the $d_{3z^2\!-\!r^2}$ orbital, giving rise to a localized spin-$\tfrac{1}{2}$ moment, while the remaining electrons 
form the broad conduction bands of $d_{x^2\!-\!y^2}$ symmetry. 

\begin{figure}
\begin{center}
\includegraphics[width=8.5cm]{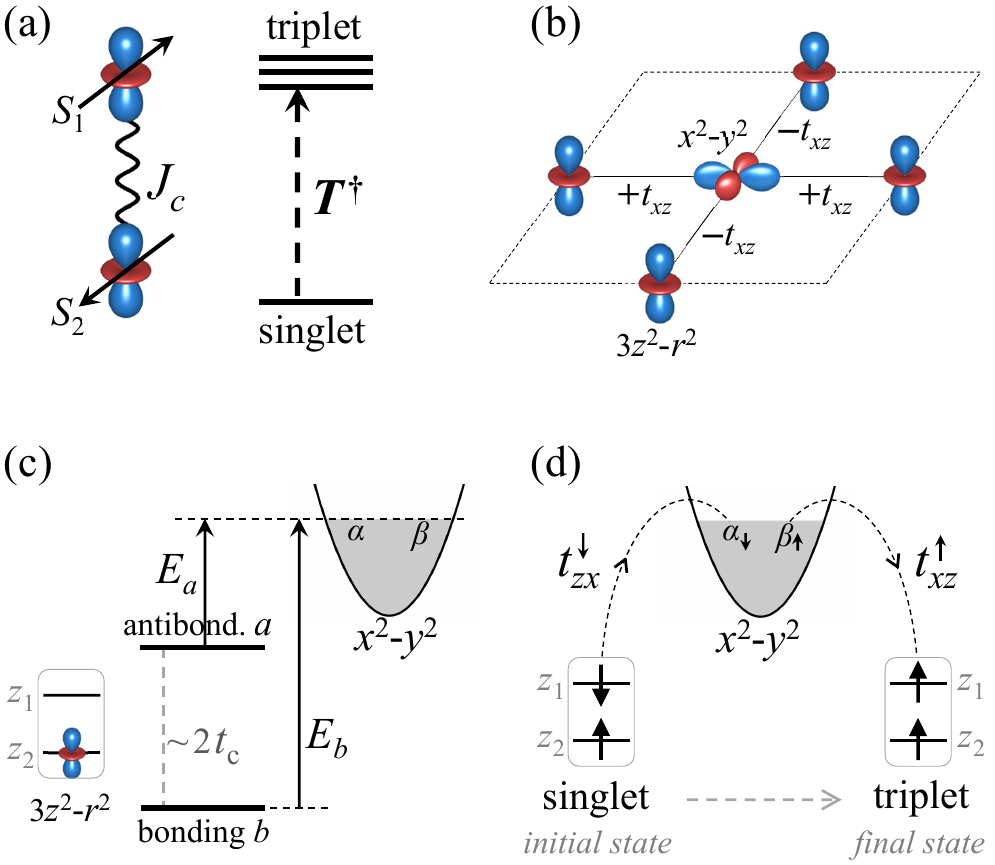}
\caption{ (a) On the $c$ axis bond connecting two layers 1 and 2, the localized $d_{3z^2\!-\!r^2}$ spins $S_1$ and $S_2$ couple 
antiferromagnetically and form a spin-singlet dimer. Singlet-to-triplet excitation (dashed arrow) with energy $J_c=4t_c^2/U$ is described by a boson creation operator $\vc T^{\dag}$. 
(b) In-plane hopping geometry between $d_{x^2\!-\!y^2}$ and $d_{3z^2\!-\!r^2}$ orbitals. The opposite signs of $\pm t_{xz}$ for $x$ and $y$ bonds generates $\eta_{\vc k}\!=\!\cos k_x \!-\!\cos k_y$ factor in Eq.~\eqref{eq:Hzx1}. 
(c) In the virtual hopping process $d_{3z^2\!-\!r^2} \!\rightarrow\! d_{x^2\!-\!y^2}$,  the $\langle z_1;z_2 \rangle$ dimer is left with a single $d_{3z^2\!-\!r^2}$ electron in the intermediate state, forming bonding $b$ and antibonding $a$ states with energies $E_b$ and $E_a$ below the Fermi level. Owing to the different excitation energies in the bonding $b\!\rightarrow\!\beta$ and antibonding $a\!\rightarrow\!\alpha$ hopping channels, the initially degenerate $\beta$ and $\alpha$ bands obtain different self-energy corrections and will split.
(d) Illustration of the Kondo exchange process where virtual $t_{zx}$ hoppings generate on-site singlet-triplet transition (triplon $T^\dagger_1$ is created in the case shown).  
}
\label{fig:1}
\end{center}
\end{figure}

Large overlap of the $d_{3z^2\!-\!r^2}$ orbitals on the $c$ axis bonds results in a strong interlayer spin interaction $\mathcal{H}_c = J_c (\vc S_1 \cdot \vc S_2)$  with $J_c = 4t_c^2/U$, where $t_c \sim 0.6$~eV and $U\sim 5$~eV are the hopping integral and intraionic Coulomb repulsion, respectively. This stabilizes interlayer spin singlets~\cite{Kha26,DMFT}, with the ground state separated from the triplet level by a gap of $J_c\sim 0.3$~eV [Fig.~\ref{fig:1}(a)]. Singlet-to-triplet spin excitations---dubbed "triplons"---are nearly local objects because the $ab$-plane hopping of $d_{3z^2\!-\!r^2}$ orbitals $t_{ab}\sim t_c/4$ and the corresponding exchange coupling $J_{ab}\sim (t_{ab}/t_c)^2 J_c$ are very small. In essence, the local spins $S=1/2$ hosted by $d_{3z^2\!-\!r^2}$ orbitals form a bilayer antiferromagnet in its singlet phase, and the spin excitations are described by hard-core vector bosons $\vc T$ with weakly dispersive energy relation $\omega_{\vc q}\simeq J_c + J_{ab}(\cos q_x +\cos q_y)\sim J_c$~\cite{Kha26}.  

 Next we address the conduction bands in a bilayer structure. While direct overlap of the planar $d_{x^2\!-\!y^2}$ orbitals on the $c$ axis bonds is negligible, there are a number of indirect interlayer hopping paths via other orbital states. A natural channel in the present case is the indirect hopping via $d_{3z^2\!-\!r^2}$ states as we consider now.   
 
Using shorthand notations $z$ and $x$ for the localized $d_{3z^2\!-\!r^2}$ and itinerant $d_{x^2\!-\!y^2}$ orbitals, respectively, the interorbital $t_{xz}$ hopping Hamiltonian is written as 
\begin{align}
\mathcal{H}_{xz} =& \sum_{i,\delta}  (\pm)_\delta \; t_{xz} \;  (x_{1,i}^{\dag} z_{1,i+\delta}+x_{2,i}^{\dag} z_{2,i+\delta}+ H.c.),
\label{eq:Hzx}
\end{align}
where the $\pm$ sign reflects the opposite phases of the $d_{x^2\!-\!y^2}$ orbital along the  $\delta=x$ and $y$ directions [see Fig.~\ref{fig:1}(b)], and the index $1,2$ labels two layers.

We introduce bonding $\beta$ and antibonding $\alpha$ conduction bands $\beta/\alpha=\tfrac{1}{\sqrt{2}}(x_1 \pm x_2)$, and similarly, the interlayer $d_{3z^2\!-\!r^2}$ molecular orbitals $b/a=\tfrac{1}{\sqrt{2}}(z_1 \pm z_2)$. In this basis, Eq.~\eqref{eq:Hzx} reads in momentum space as
\begin{align}
\mathcal{H}_{xz} =& 2 t_{xz} \sum_{\vc k}  \eta_{\vc k} \;  (\beta_{\vc k}^{\dag} b_{\vc k}+\alpha_{\vc k}^{\dag} a_{\vc k}+ H.c.),
\label{eq:Hzx1}
\end{align}
where $\eta_{\vc k}=\cos k_x - \cos k_y$. This Hamiltonian describes hybridization between localized $z$ and band $x$ states in the even and odd parity channels (with respect to the bilayer mirror symmetry). The essential point is that the excitation energies in these two channels greatly differ, due to large interlayer splitting between bonding $b$ and antibonding $a$ molecular orbitals of the order of $2t_c\sim 1.2$~eV, as illustrated in Fig.~\ref{fig:1}(c). While removing an electron from a singly-occupied $a$ orbital costs energy $E_a$, the excitation energy in the bonding channel $b\!\rightarrow\!\beta$ is much larger, $E_b\sim E_a+2t_c$. This results in different 
energy corrections $\delta \xi_{\vc k}$ to the $\alpha$ and $\beta$ band dispersions:   
\begin{align}
	\delta \xi_{\alpha\vc k} =  \epsilon_{\alpha} \eta_{\vc k}^2  \; \quad\text{with}\quad \; 
	\epsilon_{\alpha}  =4t_{xz}^2 \big(\frac{1}{E_a } -\frac{1}{E_a +U} \big), 
	\notag \\
	\delta \xi_{\beta\vc k} = \epsilon_{\beta} \eta_{\vc k}^2  \; \quad\text{with}\quad \; 
	\epsilon_{\beta}  =4t_{xz}^2 \big(\frac{1}{E_b } -\frac{1}{E_b +U} \big). 
	\label{eq:cor}
\end{align} 
Here, $E_{a/b}+U$ is the energy required to add an electron to a singly-occupied $d_{3z^2\!-\!r^2}$ level; as $U\sim 5$~eV is large, this process is less important and not shown in Fig.~\ref{fig:1}. 

\begin{figure}
	\begin{center}
		\includegraphics[width=8.5cm]{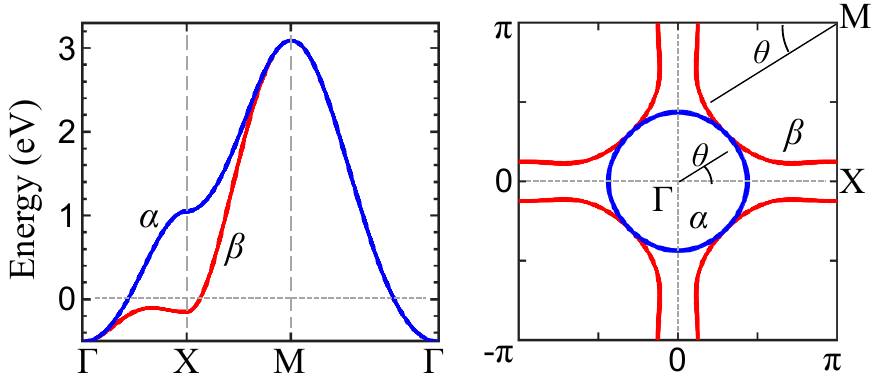}
		\caption{(left) Antibonding $\alpha$ (blue) and bonding $\beta$ (red) band dispersions along $\Gamma(0,0)$---$\mathrm{X}(\pi,0)$---$\mathrm{M}(\pi,\pi)$ path; 
		(right) the corresponding Fermi surfaces contain $n_{\alpha}+n_{\beta}=0.45$ electrons per Ni ion (as reported in ARPES study~\cite{Sun25b}). The model parameters are $t_1=0.45$, $t_2 =-0.23$, $t_3=0.07$, $\epsilon_{\alpha}=0.4$, $\epsilon_{\beta}=0.1$, and $\mu=-0.65$ (in units of eV). 
		}
		\label{fig:2}
	\end{center}
\end{figure}

The Hamiltonian for conduction bands becomes 
\begin{align}
	\mathcal {H}_c =\sum_{\vc k} (\xi_{\alpha\vc k} \; \alpha_{\vc k}^\dag \alpha_{\vc k} \;+\; \xi_{\beta\vc k} \;  \beta_{\vc k}^\dag \beta_{\vc k}),
	\label{eq:TB}
\end{align}
with the dispersion relations 
\begin{align}
	\xi_{\alpha/\beta, \vc k}=- 4 (t_1 \gamma_{1\vc k}+t_2 \gamma_{2\vc k}+t_3 \gamma_{3\vc k}) +\epsilon_{\alpha/\beta} \eta_{\vc k}^2  - \mu\;. 
	\label{eq:xi}
\end{align}
 Here $t_{1,2,3}$ are the in-plane tight-binding hopping parameters up to the third neighbors, $\gamma_{1 \vc k}=\tfrac{1}{2} ( \cos k_x + \cos k_y )$, $ \gamma_{2 \vc k}= \cos  k_x  \cos  k_y $, $ \gamma_{3 \vc k}=\tfrac{1}{2} (\cos 2k_x + \cos 2k_y)$,  and $\mu$ is the chemical potential. As  argued above, $\epsilon_{\alpha}>\epsilon_{\beta}$, thus the initial degeneracy of two bands is lifted. 

One may consider also the other virtual processes,  such as the $4s$-orbital admixtures that would contribute to $\xi_{\vc k}$ with the same $\eta_{\vc k}^2$ factor, longer-range interlayer hoppings, etc. Here we use Eq.~\eqref{eq:xi} considering $\epsilon_{\alpha/\beta}$ as free parameters. The representative $\alpha$ and $\beta$ bands and Fermi surfaces shown in Fig.~\ref{fig:2} are consistent with the ARPES data in bilayer nickelate films~\cite{Sun25b,Wan25}.

The orbital content of the $\alpha$ and $\beta$ bands is of interest for interpretation of the experiments like STM (see below). 
The above perturbative theory gives an average fraction of $d_{3z^2\!-\!r^2}$ orbitals at the Fermi energy as $p^z =\langle\nu^2 \eta_{\vc k}^2/( 1\!+\!\nu^2 \eta_{\vc k}^2)\rangle_{FS}$, where $\nu_{\alpha/\beta} = \epsilon_{\alpha/\beta}/2t_{xz}$. 
With $\epsilon_{\alpha}=0.4$~eV, $\epsilon_{\beta}=0.1$~eV as in Fig.~\ref{fig:2}, and $t_{xz}=\tfrac{\sqrt 3}{4}t_c\simeq 0.26$~eV, we find $p^z_\alpha\simeq 0.14$ and $p^z_\beta\simeq 0.054$. Note that the $\alpha$ band contains more admixture of $d_{3z^2\!-\!r^2}$ states; this can be verified by ARPES studies. 

Up to now, we described the basic ingredients of the model: triplons $\vc T$ with energy $\omega_{\vc q}\sim J_c$, and conduction bands $\alpha_{\vc k}$ and $\beta_{\vc k}$ with dispersions Eq.~\eqref{eq:xi}. Next, we derive the interactions between these degrees of freedom.   
We consider the following spin-exchange mechanisms:

(i) The intraionic Hund's coupling between local $\vc S$ and itinerant $\vc s$ spins. In a bilayer system, it reads as:
\begin{align}
	\mathcal{H}_{\mathrm H } =-2J_{\mathrm {H} } \sum_i \big[ ( \vc S \cdot \vc s)_1+ ( \vc S \cdot \vc s)_2 \big] _i. 
	\label{eq:Hund12}
\end{align}
We express local spins via triplon ${\vc T}$ operators~\cite{Sac90,Som01}: 
\begin{equation}\label{eq:T}
	{\vc S}_{1,2}=\pm\frac{1}{2}\bigl({\vc T}^\dagger+{\vc T}\bigr) 
	-\frac{i}{2}\bigl({\vc T}^\dagger\times{\vc T}\bigr), 
\end{equation}
where upper/lower sign refers to the first/second layer. In a singlet phase, where the triplons are gapped and thus their density is small, we may keep only $\vc T$-linear terms. It follows that triplons couple to the odd-parity combination of band spins $(\vc s_1 - \vc s_2)\!=\! (\vc s_{\alpha \beta} + \vc s_{\beta \alpha})$:  
\begin{align}
	\mathcal{H}_{\mathrm H } =-J_{\mathrm {H} } \sum_{\vc k  \vc k'}  (\vc T_{-\vc q}^{\dag} + \vc T_{\vc q})\cdot 
	(\vc s_{\alpha \beta}^{\vc k' \vc k} + \vc s_{\beta \alpha}^{\vc k' \vc k}). 
	\label{eq:Hund}
\end{align}
Here, the interband spin density $\vc s_{\alpha \beta}^{\vc k' \vc k} = \frac{1}{2} \alpha^{\dag}_{\vc k'  s'} {\vc \sigma}_{s's} \beta_{\vc k s}$, ${\vc \sigma}$ are the Pauli matrices, and $\vc q = \vc k'-\vc k$. 

(ii) The antiferromagnetic Kondo coupling generated by virtual hoppings between local states and conduction bands~\cite{Col15}. As illustrated in Fig.~\ref{fig:1}(d), the $t_{zx}$ hopping processes create a triplon; this is accompanied by spin-orbital flip $\alpha^\dagger_\downarrow \beta_\uparrow$ in the band sector, as required by spin-and-parity symmetry. We obtain 
\begin{align}
	\mathcal {H}_{\mathrm {K}} = \sum_{\vc k  \vc k'} \big[   \eta_{\vc k} \eta_{\vc k'} 
	(J_{\mathrm K}^a \vc T_{-\vc q}^{\dag} + J_{\mathrm K}^b \vc T_{\vc q}) \vc s_{\alpha \beta}^{ \vc k'\vc k } +H.c.\big] , 
	\label{eq:Kondo}
\end{align}
where two different exchange constants
\begin{align}
	J_{\mathrm K}^a &=   4 t_{xz}^2 \big(\frac{1}{E_a} + \frac{1}{E_a +U} \big), 
	\notag\\
	J_{\mathrm K}^b &=   4 t_{xz}^2 \big(\frac{1}{E_b} + \frac{1}{E_b +U} \big) 
	\label{eq:JK}
\end{align} 
appear due to bonding-antibonding splitting of the $d_{3z^2\!-\!r^2}$ levels in the intermediate state, i.e., for the same reason as in the density channel above, cf. Eq.~\eqref{eq:cor}.     

The total interaction $\mathcal {H}_{\mathrm {K} } + \mathcal {H}_{\mathrm {H} }$ between singlet-triplet excitations $\vc T$ and itinerant spins reads then as 
\begin{align} 
\mathcal {H}_{\mathrm {int}} = - J_{\mathrm H} \sum_{\vc k  \vc k'}  
\big[(f_{\vc k\vc k'}^a \vc T_{-\vc q}^{\dag} +  f_{\vc k\vc k'}^b \vc T_{\vc q}) \vc s_{\alpha \beta}^{ \vc k' \vc k}  + H.c. \big],
\label{eq:int}
\end{align}
where $ f_{\vc k\vc k'}^{a/b}=(1\!-\!\kappa_{a/b} \eta_{\vc k} \eta_{\vc k'})$. 
The parameters $\kappa_a=J_{\mathrm K}^a/J_{\mathrm H}$ and $\kappa_b=J_{\mathrm K}^b/J_{\mathrm H}$ quantify the intersite Kondo coupling relative to intraionic $J_H$. The $\kappa$-terms render the interaction nonlocal and, as shown below, lead to the anisotropy of the superconducting gaps. 

Few comments are in order here. We assume that the ferromagnetic Hund's coupling is strong enough to prevent Kondo instability and related heavy fermion physics. On the other hand, we assume that Hund's splitting $2J_{\mathrm H}$ in nickelates is well below the $e_g$ electron bandwidth of $W\sim3-4$~eV, i.e., it is not strong enough to form $S=1$ local spins and stabilize their magnetic ordering by virtue of the double exchange mechanism. Under these assumptions, we will treat $\mathcal {H}_{\mathrm {int}}$ of  Eq.~\eqref{eq:int} perturbatively. 
 
 Finally, in contrast to the Stoner continuum and paramagnons that mediate pairing in standard spin-fluctuation models, the triplon excitations $\vc T$ are quasi-local, weakly-dispersive modes with the unusually large energy scale $J_c\!\sim \!0.3$~eV arising from the strong interlayer overlap of the $d_{3z^2\!-\!r^2}$ orbitals. Experimentally, they can be detected by inelastic neutron and resonant x-ray scattering, with the largest intensity at $q_z=\pi$.        
 
  {\it Cooper pairing and the  gap structure}.---The exchange of triplet excitations with energy $\sim J_c$ generates an effective interaction between electrons~\cite{Kha26}. Treating Eq.~\eqref{eq:int} perturbatively, we obtain the pairing Hamiltonian
\begin{align}
\mathcal{H}_{\rm BCS}= \sum_{\vc k\vc k'} V_{\vc k\vc k'} (\alpha^{\dag}_{\vc k \uparrow}\alpha^{\dag}_{-\vc k \downarrow} \beta_{-\vc k' \downarrow} \beta_{\vc k' \uparrow} +H.c.) , 
\label{eq:BCS}
\end{align}
with the momentum dependent matrix element 
\begin{align}
V_{\vc k\vc k'} =V f_{\vc k\vc k'}^a f_{\vc k\vc k'}^b = V(1-\kappa_a \eta_{\vc k} \eta_{\vc k'})(1-\kappa_b \eta_{\vc k} \eta_{\vc k'}) .
\label{eq:V}
\end{align}
The coupling constant $V=3J_{\mathrm H}^2/2 J_c$. Here $J_{\mathrm H}$ and $ J_c$ play roles analogous to the electron-phonon coupling strength and phonon energy, respectively, in the phonon-mediated pairing theory.  

BCS mean-field treatment of the Hamiltonian~\eqref{eq:BCS} leads to the following equations for the pairing gaps $\Delta_{ \alpha \vc k}$ and $\Delta_{ \beta \vc k}$ on the $\alpha$ and $\beta$ Fermi-surfaces:
\begin{align}
\Delta_{\alpha\vc k}&=- \sum_{\vc k'} V_{\vc k\vc k'} \;\frac{\Delta_{\beta\vc k'}}{2\varepsilon_{\beta \vc k'}}  \tanh \frac{\varepsilon_{\beta \vc k'}}{2T}\;,
\notag \\
\Delta_{\beta\vc k}&=- \sum_{\vc k'} V_{\vc k\vc k'} \;\frac{\Delta_{\alpha\vc k'}}{2\varepsilon_{\alpha \vc k'}}  \tanh \frac{\varepsilon_{\alpha \vc k'}}{2T}\;.
\label{eq:gap1}
\end{align}
The quasiparticle energies 
$\varepsilon_{\alpha \vc  k} = \sqrt{\xi_{\alpha \vc k}^2+\Delta_{ \alpha \vc k}^2}$ and 
$\varepsilon_{\beta \vc  k} = \sqrt{\xi_{\beta \vc k}^2+\Delta_{\beta \vc k}^2}$, with $\xi_{ \alpha/\beta}$ from Eq.~\eqref{eq:xi}. 
The two gap equations are coupled, reflecting the interband pairing nature of the interactions in Eq.~\eqref{eq:BCS}. 

The leading solution of Eq.~\eqref{eq:gap1} corresponds to a so-called $s_{\pm}$-wave state, in which the gaps on the $\alpha$ and $\beta$ bands have opposite signs, i.e., ${\rm sgn}(\Delta_{\alpha})\!=\!-{\rm sgn}(\Delta_{\beta})$. Henceforth, we omit the gap signs and take $\Delta_{\alpha/\beta}$ to denote the gap magnitudes, $\Delta_{\alpha/\beta} \!\equiv\!|\Delta_{\alpha/\beta}|$. For $s$-wave symmetry, the $\kappa$-linear terms in $V_{\vc k\vc k'}$ vanish upon momentum summation in Eq.~\eqref{eq:gap1}, yielding the gap equations at $T=0$:
\begin{align}
	\Delta_{\alpha\vc k}&=V\sum_{\vc p} (1 + \kappa^2 \eta_{\vc k}^2 \eta_{\vc p}^2) \;\frac{\Delta_{\beta\vc p}}{2\varepsilon_{\beta \vc p}}  \;,
	\notag \\
	\Delta_{\beta\vc k}&=V\sum_{\vc p} (1 + \kappa^2 \eta_{\vc k}^2 \eta_{\vc p}^2) \;\frac{\Delta_{\alpha\vc p}}{2\varepsilon_{\alpha \vc p}}  \;,
	\label{eq:gap2}
\end{align}
where $\kappa^2=\kappa_a \kappa_b=J_{\mathrm K}^a J_{\mathrm K}^b/J_{\mathrm H}^2$. 

\begin{figure}
\begin{center}
\includegraphics[width=8.5cm]{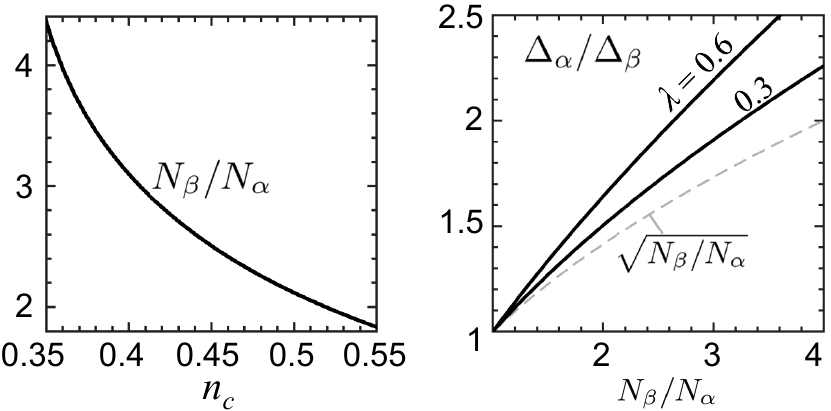}
\caption{
(left) Ratio $N_{\beta}/N_{\alpha}$ of the DOS for $\alpha$ and $\beta$ bands of Fig.~\ref{fig:2} as a function of electron density $n_c = n_\alpha + n_\beta$ per Ni. 
(right) Ratio of the superconducting gap amplitudes $\Delta_\alpha/\Delta_\beta$ as a function of $N_{\beta}/N_{\alpha}$, following from Eq.~\eqref{eq:Dab} at $\lambda=0.6$ and $0.3$.  Dashed line shows $\lambda \rightarrow 0$ result  $\Delta_\alpha/\Delta_\beta=\sqrt{N_{\beta}/N_{\alpha}}$.
}
\label{fig:3}
\end{center}
\end{figure}

We approximate the momentum summations by integrals (with energy cutoff $\Omega \! \sim \! J_c$): 
$\sum_{\vc p} (\Delta_{\zeta\vc p}/2\varepsilon_{\zeta\vc p}) \! \approx \! N_\zeta\Delta_\zeta\ln\frac{2\Omega}{\Delta_\zeta}$ and 
$\sum_{\vc p} \eta_{\vc p}^2(\Delta_{\zeta\vc p}/2\varepsilon_{\zeta\vc p}) \! \approx \! \widetilde{N}_\zeta\Delta_\zeta\ln\frac{2\Omega}{\Delta_\zeta}$, where $N_\zeta$ is the density of states (DOS) per spin of the band $\zeta \! \in \! \{\alpha, \beta \}$, and $\widetilde{N}_\zeta$ denotes the DOS weighted by $\eta_{\vc k}^2$. This yields the following momentum dependence of the gaps: 
   \begin{align}
  \Delta_{\alpha\vc k}&= \Delta_\alpha  \bigg[1+ \frac{\widetilde{N}_\beta}{N_\beta}  \kappa^2 \eta_{\vc k}^2 \bigg], 	
  \notag \\
  \Delta_{\beta\vc k} &= \Delta_\beta  \bigg[1+ \frac{\widetilde{N}_\alpha}{N_\alpha}  \kappa^2 \eta_{\vc k}^2 \bigg].
  \label{eq:gap3}
  \end{align}
 The gap values for the diagonal $\theta=\pi/4$ direction $\Delta_\alpha$ and $\Delta_\beta$ are given by the coupled equations 
  \begin{align}
  	\Delta_\alpha= \! VN_\beta\Delta_\beta\ln\frac{2\Omega}{\Delta_\beta} \quad\text{and}\quad \!
  	\Delta_\beta = \! VN_\alpha\Delta_\alpha\ln\frac{2\Omega}{\Delta_\alpha}, 
  	\label{eq:gap4}
  \end{align}
 from which we obtain 
 \begin{align}
 	\Delta_\alpha \! \simeq \Delta \big(N_\beta/N_\alpha\big)^{\textstyle\frac{1}{2(2-\lambda)}}, \; 
 	\Delta_\beta \! \simeq \Delta \big(N_\alpha/N_\beta\big)^{\textstyle\frac{1}{2(2-\lambda)}}, 
 	\label{eq:Dab}
 \end{align}
 where $\Delta\simeq 2\Omega\exp(-1/\lambda)$ and $\lambda=V\sqrt{N_\alpha N_{\beta}}=\sqrt{\lambda_\alpha\lambda_\beta}$. This approximation is valid if $\lambda \ln\frac{\Delta_\alpha}{\Delta_\beta} < 2$.
 
\begin{figure}
	\begin{center}
		\includegraphics[width=8.6cm]{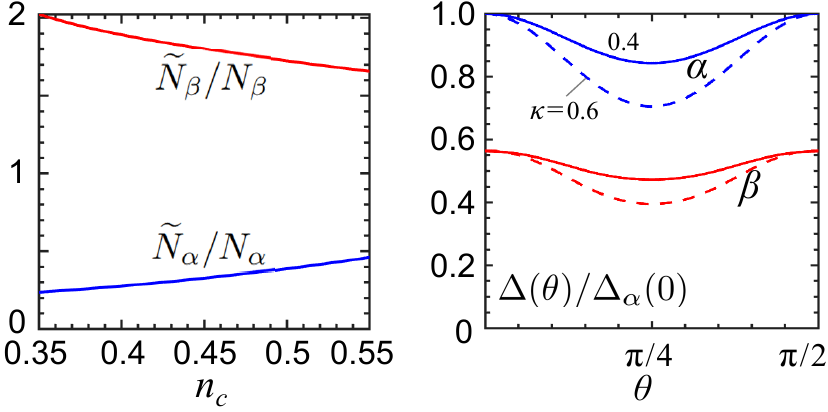}
		\caption{ 
			(left) The $\eta_{\vc k}^2$-weighted DOS $\widetilde{N}$ (relative to DOS $N$) as a function of electron density $n_c$.  
			(right) The gap values [relative to $\Delta_\alpha(\theta\!=\!0$)] on the $\alpha$ and $\beta$ Fermi surfaces as a function of angle $\theta$ (defined in Fig.~2), calculated from Eqs.~\eqref{eq:gap3} and~\eqref{eq:gap4} with $n_c=0.45$, $\lambda=0.4$, and two different $\kappa$ values. 
		}
		\label{fig:4}
	\end{center}
\end{figure}

The ratio $N_{\beta}/N_{\alpha}$ depends on the band structure parameters, and also on the electron density $n_c$ (Fig.~\ref{fig:3}). The evolution of $N_{\beta}/ N_{\alpha} $ strongly impacts the gap hierarchy, leading to large changes in $\Delta_\alpha / \Delta_\beta$ ratio, see Fig.~\ref{fig:3}(right). Remarkably, the larger gap emerges on the $\alpha$-sheet despite its smaller DOS at the Fermi level. This counterintuitive observation originates from the interband nature of the triplon-mediated pairing. 

According to Eq.~\eqref{eq:gap3}, anisotropy of the superconducting gap $\Delta_{\alpha \vc k}$  on the $\alpha$ band depends on the ratio $\widetilde{N}_\beta/ N_\beta$ of the ``weighted" to ``bare" density of states on the $\beta$-sheet, and vice versa. The $\widetilde{N}/ N$ values depend on the band filling (Fig.~\ref{fig:4}). We recall that the gap anisotropy is due to nonlocal Kondo coupling, represented by the $\kappa$ terms in the pairing potential $V_{\vc k\vc k'}$~\eqref{eq:V}. 

Figure~\ref{fig:4}(right) shows the gap structure calculated using the bands and Fermi surfaces of Fig.~\ref{fig:2}. While quantitative details depend on the parameters used, the key features, such as the gap hierarchy $\Delta_\alpha > \Delta_\beta$ and the angular dependencies, are robust. These findings are in contrast to the predictions of conventional spin-fluctuation and $t\!-\!J$ models, which report the opposite relation $\Delta_\alpha < \Delta_\beta$ and nearly vanishing gaps at certain momenta (see, e.g., Refs.~\cite{Qiu25,Wat26}). 

\begin{figure}
\begin{center}
\includegraphics[width=8.5cm]{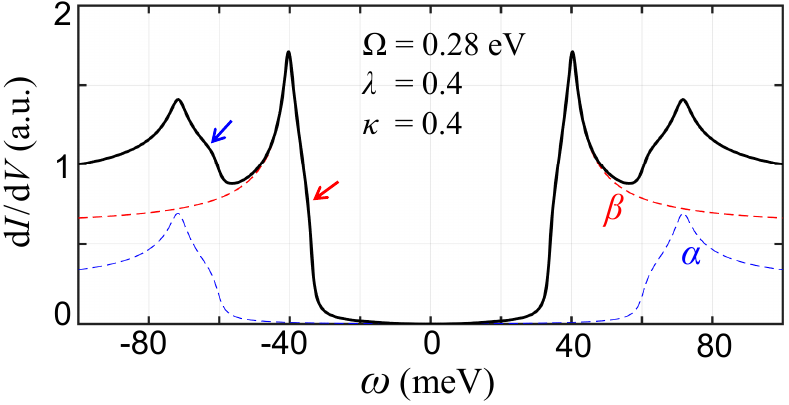}
\caption{Tunneling spectrum is dominated by the $\beta$ band (red line) owing to its larger DOS $N_\beta$. The weak shoulders of the peaks (indicated by arrows) are due to the gap anisotropy arising from Kondo coupling $\kappa$. Parameters $p_{\alpha}^z=0.14$, $p_{\beta}^z=0.055$ were calculated using the bands of Fig.~\ref{fig:2} (see text), and $r=0.7$ is used. The broadening $\Gamma_{\beta}=1$ and $\Gamma_{\alpha}=2$~meV.
}
\label{fig:5}
\end{center}
\end{figure}

As a direct test of our theory, we have calculated the tunneling spectra. Following Ref.~\cite{Dyn84}, we express zero temperature 
differential conductance via a broadened and angular averaged BCS density of states $N(\omega)=N Re\langle(\omega-i\Gamma)/[(\omega-i\Gamma)^2-\Delta^2]^{1/2}\rangle_\theta$ for $\alpha$ and $\beta$ bands: 
$dI/dV \propto (p^z_\alpha\!+\!r p^x_\alpha) N_{\alpha}(\omega) + (p^z_\beta\!+\!r p^x_\beta) N_{\beta}(\omega)$. The prefactors depend on the  orbital content $p^z$ and $p^x\!=\!1\!-\!p^z$ of the bands, while parameter $r\!<\!1$ accounts for the orbital selectivity of the experimental probes like STM. For a broad range of $p^z$ and $r$, the $\beta$ band with large DOS $N_{\beta}$ most contributes to the spectra. 

A representative spectrum in Fig.~\ref{fig:5} shows the leading $\beta$ peak at $40$~meV and less intense $\alpha$ peak around $70$~meV, reproducing the two distinct peaks observed in recent STM spectra~\cite{Lia26}. Their positions are determined by two parameters, $\Omega\!\sim\! J_c$ and $\lambda$, while the lineshapes depend on the anisotropy parameter $\kappa$. Remarkably, their values 
in Fig.~\ref{fig:5} agree well with the predictions of the triplon-mediated pairing model~\cite{Kha26}. The large pairing gaps observed experimentally~\cite{Lia26}, together with their quantitative description by the present theory, suggest that substantially higher transition temperatures may be achievable in bilayer nickelates once high-quality, homogeneous samples become available~\cite{noteTc}. 

In summary, we have developed a microscopic theory of the superconducting gap structure in bilayer nickelates, based on a model in which itinerant $d_{x^2-y^2}$ electrons interact with localized $d_{3z^2-r^2}$ spins that form interlayer spin-singlet dimers. Including orbital hybridization effects, we derived the effective bands and Fermi surfaces consistent with ARPES data, and derived the interactions between local and conduction electron spins. 

The triplon excitations of the spin-dimer background mediate Cooper pairing of the $s_{\pm}$ symmetry, with opposite signs of the order parameter on the $\alpha$ and $\beta$ bands. The nonlocal Kondo exchange induces a sizable anisotropy in the pairing gaps. The results are presented as analytical expressions, which facilitate their comparison with experimental data. 

The theory accounts for the key experimental observations~\cite{Fan25,Lia26}. The $\alpha$ band hosts a larger superconducting gap than the $\beta$ band despite its smaller density of states. This hierarchy follows from the interband nature of triplon-mediated pairing, where the gap amplitude on one Fermi surface is controlled by the density of states on the other. The gaps also exhibit angular anisotropy with a minimum along the $k_x=k_y$ direction. Taken together, these findings provide strong support for the triplon-mediated pairing mechanism, helping to resolve the ongoing debate on superconductivity in nickelates.  

We thank J.~Chaloupka, M.~Hepting, I.~M. Eremin, H.-H. Wen, J. Wang, and B.~Keimer for useful discussions.  
H.~L.  is supported by the National Natural Science Foundation of China under the Grant No. 12574066 and by the Fundamental Research Funds for the Central Universities under the Grant No. KG202501. G.~Kh. acknowledges support from the European Research Council under Advanced Grant No.~101141844 (SpecTera). 

{\it Data availability}.---The data supporting the findings of this study are available from the corresponding author upon reasonable request.

\bibliographystyle{apsrev4-2}

\begin{thebibliography}{99}

\bibitem{Sun23a}
H. Sun, M. Huo, X. Hu, J. Li, Z. Liu, Y. Han, L. Tang, Z. Mao, P. Yang, B. Wang, J. Cheng, D.-X. Yao, G.-M. Zhang, and M. Wang,
 Signatures of superconductivity near 80 K in a nickelate under high pressure,
Nature \textbf{621}, 493 (2023).

\bibitem{Wan24}
G. Wang, N. N. Wang, X. L. Shen, J. Hou, L. Ma, L. F. Shi, Z. A. Ren, Y. D. Gu, H. M. Ma, P. T.Yang, Z. Y. Liu, H. Z. Guo, J. P. Sun, G. M. Zhang,
S. Calder, J.-Q. Yan, B. S. Wang, Y. Uwatoko, and J.-G. Cheng, 
Pressure-induced superconductivity in polycrystalline La$_3$Ni$_2$O$_{7-\delta}$, Phys. Rev. X \textbf{14}, 011040 (2024).

\bibitem{Ko25}
E. K. Ko, Y. Yu, Y. Liu, L. Bhatt, J. Li, V. Thampy, C.-T. Kuo, B. Y. Wang, Y. Lee, K. Lee, J.-S. Lee, B. H. Goodge, D. A. Muller, and H. Y. Hwang,
 Signatures of ambient pressure superconductivity in thin film La$_3$Ni$_2$O$_7$,
Nature \textbf{638}, 935 (2025).

\bibitem{Zho25}
G. Zhou, W. Lv, H. Wang, Z. Nie, Y. Chen, Y. Li, H. Huang, W.-Q. Chen, Y.-J. Sun, Q.-K. Xue, and Z. Chen,
 Ambient-pressure superconductivity onset above 40 K in (La,Pr)$_3$Ni$_2$O$_7$ films,
Nature \textbf{640}, 641 (2025).

\bibitem{Liu25}
Y. Liu, E. K. Ko, Y. Tarn, L. Bhatt, B. H. Goodge, D. A. Muller, S. Raghu, Y. Yu, and H. Y. Hwang,
 Superconductivity and normal-state transport in compressively strained La$_2$PrNi$_2$O$_7$ thin films,
arXiv:2501.08022.

\bibitem{Pup25}
P. Puphal, T. Sch\"{a}fer, B. Keimer, and M. Hepting, 
Superconductivity in infinite-layer and Ruddlesden--Popper nickelates, 
Nat. Rev. Phys.  \textbf{8}, 70 (2025)

\bibitem{Luo23}
Z. Luo, X. Hu, M. Wang, W. W$\acute{\rm u}$, and D.-X. Yao, Bilayer two-orbital model of La$_3$Ni$_2$O$_7$ under pressure,
Phys. Rev. Lett. \textbf{131}, 126001 (2023).

\bibitem{Yan23a}
Q.-G. Yang, D. Wang, and Q.-H. Wang,
 Possible $s^\pm$-wave superconductivity in La$_3$Ni$_2$O$_7$,
Phys. Rev. B \textbf{108}, L140505 (2023).

\bibitem{Yan23b}
Y.-F. Yang, G.-M. Zhang, and F.-C. Zhang,
Interlayer valence bonds and two-component theory for high-$T_c$ superconductivity of La$_3$Ni$_2$O$_7$ under pressure,
Phys. Rev. B \textbf{108}, L201108 (2023).

\bibitem{Qin23}
Q. Qin and Y.-F. Yang,
High-$T_c$ superconductivity by mobilizing local spin singlets and possible route to higher $T_c$ in pressurized La$_3$Ni$_2$O$_7$,
Phys. Rev. B \textbf{108}, L140504 (2023).

\bibitem{Liu23}
Y.-B. Liu, J.-W. Mei, F. Ye, W.-Q. Chen, and F. Yang,
 $s^\pm$-wave pairing and the destructive role of apical-oxygen deficiencies in La$_3$Ni$_2$O$_7$ under pressure,
Phys. Rev. Lett.  \textbf{131}, 236002 (2023).

\bibitem{Oh23}
H. Oh and Y.-H. Zhang, 
Type-II $t-J$ model and shared superexchange coupling from Hund's rule in superconducting La$_3$Ni$_2$O$_7$, 
Phys. Rev. B \textbf{108}, 174511 (2023).
 
 \bibitem{Lia23}
 Z. Liao, L. Chen, G. Duan, Y. Wang, C. Liu, R. Yu, and Q. Si, 
 Electron correlations and superconductivity in La$_3$Ni$_2$O$_7$ under pressure tuning, 
 Phys. Rev. B \textbf{108}, 214522 (2023).

\bibitem{Lec23}
F. Lechermann, J. Gondolf, S. B\'otzel, and I. M. Eremin,
Electronic correlations and superconducting instability in La$_3$Ni$_2$O$_7$ under high pressure,
Phys. Rev. B \textbf{108}, L201121 (2023).

\bibitem{Qu23}
X.-Z. Qu, D.-W. Qu, X.-W. Yi, W. Li, G. Su, 
Hund's rule, interorbital hybridization, and high-$T_c$ superconductivity in the bilayer nickelate, arXiv:2311.12769. 

\bibitem{Jia25}
K.-Y. Jiang, Y.-H. Cao, Q.-G. Yang, H.-Y. Lu, and Q.-H. Wang,
Theory of pressure dependence of superconductivity in bilayer nickelate La$_3$Ni$_2$O$_7$,
Phys. Rev. Lett. \textbf{134}, 076001 (2025).

\bibitem{Sak24}
H. Sakakibara, N. Kitamine, M. Ochi, and K. Kuroki,
Possible high-$T_c$ superconductivity in La$_3$Ni$_2$O$_7$ under high pressure through manifestation of a nearly half-filled bilayer Hubbard model,
Phys. Rev. Lett.  \textbf{132}, 106002 (2024).

\bibitem{Che24}
J. Chen, F. Yang, and W. Li, 
Orbital-selective superconductivity in the pressurized bilayer nickelate La$_3$Ni$_2$O$_7$:
An infinite projected entangled-pair state study, 
Phys. Rev. B \textbf{110}, L041111 (2024).

\bibitem{Lu24}
C. Lu, Z. Pan, F. Yang, and C. Wu, 
Interplay of two $E_g$ orbitals in superconducting La$_3$Ni$_2$O$_7$ under pressure,
Phys. Rev. B \textbf{110}, 094509 (2024).

\bibitem{Yi25}
X.-W. Yi, W. Li, J.-Y. You, B. Gu, and G. Su, 
Unifying strain- and pressure-driven superconductivity in La$_3$Ni$_2$O$_7$: Suppressed charge and spin density waves and enhanced interlayer coupling, 
Phys. Rev. B \textbf{112}, L140504 (2025).

\bibitem{Kha26}
G. Khaliullin and J. Chaloupka, Orbital order and superconductivity in bilayer nickelate compounds,
Phys. Rev. B \textbf{113}, L041115 (2026).

\bibitem{Qiu25}
W. Qiu, Z. Luo, X. Hu, and D.-X. Yao,  
Pairing symmetry and superconductivity in La$_3$Ni$_2$O$_7$ thin films, arXiv:2506.20727. 

\bibitem{Wat26}
H. Watanabe, H. Sakakibara, and K. Kuroki, 
Hierarchical structure of primary and hybridization-induced superconducting correlations in bilayer nickelates, arXiv:2603.13604. 

\bibitem{Wang25}
J. Wang, Y.-F. Yang,
Fermi liquid and isotropic superconductivity of Hund scenario for bilayer nickelates, arXiv:2507.19301. 

\bibitem{Fan25}
S. Fan, M. Ou, M. Scholten, Q. Li, Z. Shang, Y. Wang, J. Xu, H. Yang, I. M. Eremin, and H.-H. Wen, 
Superconducting gap structure and bosonic mode in La$_2$PrNi$_2$O$_7$ thin films at ambient pressure, arXiv:2506.01788.

\bibitem{She25}
J. Shen, G. Zhou, Y. Miao, P. Li, Z. Ou, Y. Chen, Z. Wang, R. Luan, H. Sun, Z. Feng, X. Yong, Y. Li, L. Xu, W. Lv, Z. Nie, H. Wang, H. Huang, 
Y.-J. Sun, Q.-K. Xue, J. He, Z. Chen, 
Nodeless superconducting gap and electron-boson coupling in (La,Pr,Sm)$_3$Ni$_2$O$_7$ films, arXiv:2502.17831.

\bibitem{Sun25b}
W. Sun, Z. Jiang, B. Hao, S. Yan, H. Zhang, M. Wang, Y. Yang, H. Sun, Z. Liu, D. Ji, Z. Gu, J. Zhou, D. Shen, D. Feng, and Y. Nie, Observation of superconductivity-induced leading-edge gap in Sr-doped La$_3$Ni$_2$O$_7$ thin films, arXiv:2507.07409.
 
 \bibitem{Wan26}
 X. Wang, Y. Chen, C. Ding, L. Xu, J.-J. Miao, G. Zhou, Z. Chen, Y.-J. Sun, J.-F. Jia, Q.-K. Xue, 
 Atomically resolved intrinsic superconducting gap in (La,Pr)$_3$Ni$_2$O$_7$ films, arXiv:2605.14806.
 
 \bibitem{Lia26}
 Z. Liang, T. Wei, W. Ren, H. Ji, Z. Xie, Y. Liu, Z. Wang, J. Wang, 
 Observation of flat-bottom U-shaped energy gap in high-Tc nickelate (La,Pr)$_3$Ni$_2$O$_7$ thin films, arXiv:2605.15703.
 
\bibitem{Guo25}
J. Guo, Y. Chen, Y. Wang, H. Sun, D. Hu, M. Wang, X. Huang, and T. Cui,
Revealing superconducting gap in La$_3$Ni$_2$O$_{7-\delta}$ by Andreev reflection spectroscopy under high pressure, arXiv:2509.12601.

\bibitem{Cao25}
Z.-Y. Cao, D. Peng, S. Choi, F. Lan, L. Yu, E. Zhang, Z. Xing, Y. Liu, F. Zhang, T. Luo, L. Chen, V. T. A. Hong, S.-Y. Paek, H. Jang, J. Xie, H. Liu, H. Lou, Z. Zeng, Y. Ding, J. Zhao, C. Liu, T. Park, Q. Zeng, and H.-K. Mao,
Direct observation of d-wave superconducting gap symmetry in pressurized La$_3$Ni$_2$O$_{7-\delta}$ single crystals, arXiv:2509.12606.

\bibitem{Bha25}
H. C. R. B. Bhatta, X. Zhang, Y. Zhong, and C. Jia,
Structural and electronic evolution of bilayer nickelates under biaxial strain, arXiv:2502.01624.

\bibitem{DMFT} Formation of the interlayer spin singlets has recently been confirmed by cluster DMFT study of Ref.~\cite{Lab26}. 

\bibitem{Lab26} H. LaBollita, A. J. Millis, and O. Gingras, 
Squeezing dynamical singlets in bilayer nickelates, arXiv:2606.07199.

\bibitem{Wan25}
B. Y. Wang, Y. Zhong, S. Abadi, Y. Liu, Y. Yu, X. Zhang, Y.-M. Wu, R. Wang, J. Li, Y. Tarn, E. K. Ko, V. Thampy, M. Hashimoto, D. Lu, Y. S. Lee,
T. P. Devereaux, C. Jia, H. Y. Hwang, and Z.-X. Shen,
Electronic structure of compressively strained thin film La$_2$PrNi$_2$O$_7$, arXiv:2504.16372.

\bibitem{Sac90}
S. Sachdev and R. N. Bhatt, Bond-operator representation of quantum spins: Mean-field theory of frustrated
quantum Heisenberg antiferromagnets, 
Phys. Rev. B \textbf{41}, 9323 (1990).

\bibitem{Som01}
T. Sommer, M. Vojta, and K. W. Becker, 
Magnetic properties and spin waves of bilayer magnets in a uniform field, 
Eur. Phys. J. B \textbf{23}, 329 (2001).

\bibitem{Col15}
P. Coleman, {\it Introduction to Many-Body Physics} 
(Cambridge University Press, Cambridge, 2015), Chap. 16. 

\bibitem{Dyn84}
R. C. Dynes, J. P.  Garno, G. B. Hertel, and T. P. Orlando, Tunneling study of superconductivity near the metal-insulator transition, Phys. Rev. Lett. \textbf{53}, 2437 (1984).
  
 \bibitem{noteTc} Significant structural and electronic inhomogeneity revealed by local probes~\cite{Man25,Bhat25,Fan25,Wan26,Lia26} likely strongly reduce the current $T_c$ values in nickelates.  
  
  \bibitem{Man25}
  S. V. Mandyam, E. Wang, Z. Wang, B. Chen, N. C. Jayarama, A. Gupta, E. A. Riesel, V. I. Levitas, C. R. Laumann, and N. Y. Yao, 
  Uncovering origins of heterogeneous superconductivity in La$_3$Ni$_2$O$_7$ using quantum sensors, arXiv:2510.02429.
  
  \bibitem{Bhat25} 
  L. Bhatt, A. Y. Jiang, E. K. Ko, N. Schnitzer, G. A. Pan, D. F. Segedin, Y. Liu, Y. Yu, Y.-F. Zhao, E. A. Morales, C. M. Brooks,
  A. S. Botana, H. Y. Hwang, J. A. Mundy, D. A. Muller, and B. H. Goodge, 
  Resolving structural origins for superconductivity in strain-engineered La$_3$Ni$_2$O$_7$ thin films, arXiv:2501.08204.
  
\end{thebibliography}

\end{document}